\documentstyle[12pt]{article}
\begin{document}
\begin{flushright}
\baselineskip=12pt
DOE/ER/40717--29\\
CTP-TAMU-23/96\\
ACT-08/96\\
\tt hep-ph/9607220
\end{flushright}

\begin{center}
\vglue 1.5cm
{\Large\bf A supergravity explanation of the CDF $ee\gamma\gamma$ event}
\vglue 2cm
{\Large Jorge L. Lopez$^1$ and D.V. Nanopoulos$^{2,3}$}
\vglue 1.5cm
\begin{flushleft}
$^1$Department of Physics, Bonner Nuclear Lab, Rice University\\ 6100 Main
Street, Houston, TX 77005, USA\\
$^2$Center for Theoretical Physics, Department of Physics, Texas A\&M
University\\ College Station, TX 77843--4242, USA\\
$^3$Astroparticle Physics Group, Houston Advanced Research Center (HARC)\\
The Mitchell Campus, The Woodlands, TX 77381, USA\\
\end{flushleft}
\end{center}

\vglue 2cm
\begin{abstract}
We present a unified no-scale supergravity model with a light gravitino that
can naturally explain the observed $ee\gamma\gamma$ event at CDF via
right-handed selectron pair-production. The full spectrum of our model can be
described in terms of a {\em single parameter} and can be distinguished from
alternative proposals in the literature. Ongoing and future runs at LEP 2
should be able to probe the full allowed parameter space via acoplanar diphoton
events from $e^+e^-\to\chi^0_1\chi^0_1$ production.
\end{abstract}
\vspace{1.5cm}
\begin{flushleft}
\baselineskip=12pt
DOE/ER/40717--29\\
CTP-TAMU-23/96\\
ACT-08/96\\
July 1996\\
\end{flushleft}
\newpage
\setcounter{page}{1}
\pagestyle{plain}
\baselineskip=14pt

Supersymmetry is widely acknowledged to be the best motivated extension of
the Standard Model of particle physics. Yet, experimental searches over the
last decade have failed to provide direct and unambiguous evidence for the
existence of the predicted rich spectrum of superparticles, although plenty
of indirect evidence exists, such as the implied light Higgs-boson mass
($m_h=76^{+152}_{-50}\,{\rm GeV}$ \cite{EFL}) from fits to the precise
electroweak data, and the celebrated unification of the Standard Model gauge
couplings in supersymmetric grand unified theories \cite{EKN}. As it has been
recently pointed out \cite{KanePRL,DinePRL}, this trend may be coming
to an end with the observation by the CDF Collaboration \cite{Park} of one
event of the type $e^+e^-\gamma\gamma+E_T\hskip-13pt/\ $. This event has no
conceivable Standard Model explanation, but can be ascribed to supersymmetry in
two possible contexts, depending on whether the lightest supersymmetric
particle is the lightest neutralino ($\chi^0_1$) or the gravitino ($\widetilde
G$). In both scenarios it is assumed that the CDF event results from selectron
pair production. In the `neutralino LSP' scenario the selectrons must decay in
a rather cumbersome way: $\tilde e\to e\chi^0_2$, $\chi^0_2\to\chi^0_1\gamma$.
The loop-induced radiative neutralino decay may occur in small regions of
parameter space, but these regions are inconsistent with the traditional
gaugino mass unification condition. In the `gravitino LSP' scenario the
selectron decay is much more straightforward: $\tilde e\to e\chi^0_1$,
$\chi^0_1\to \gamma\widetilde G$. In neither scenario can one ascertain whether
$\tilde e=\tilde e_R$ or $\tilde e=\tilde e_L$.

Notably lacking from the above discussion is the presence of the theoretical
framework that has propelled the study of supersymmetry, namely {\em grand
unification} and {\em supergravity}. In the neutralino LSP scenario one must
seemingly abandon the GUT relation among the gaugino masses
\cite{KanePRL,Kane}, whereas the gravitino LSP scenario has been couched in the
context of low-energy gauge-mediated supersymmetry breaking
\cite{DinePRL,Dine}. In this Letter we propose an explanation for the CDF event
in the context of grand unification and supergravity. Our framework fits within
the class of no-scale supergravity models which illuminate the cosmological
constant problem \cite{no-scale}, may allow the dynamical determination of all
mass scales \cite{Lahanas}, allow a very light gravitino \cite{EEN1}, may
solve the strong CP problem \cite{EEN1,EEN2}, and may be
derivable from string \cite{LN}. Our {\em one-parameter} model thus realizes
the more appealing gravitino LSP scenario, where
$m_{3/2}\raisebox{-4pt}{$\,\stackrel{\textstyle{<}}{\sim}\,$}1\,{\rm KeV}$
has well-known cosmological advantages \cite{cosmo} and, through the
inside-the-detector decay $\chi^0_1\to\gamma\widetilde G$, makes the
ever-present lightest neutralino readily identifiable. In practice our proposal
amounts to setting the universal scalar mass and the universal trilinear scalar
couplings to zero ($m_0=A_0=0$). Our sole source of supersymmetry breaking is
the gaugino mass $m_{1/2}$, which entails a low-energy supersymmetric spectrum
described in terms of a single parameter, thus providing several correlated and
falsifiable predictions, that can be distinguished from those in the
alternative models of Refs.~\cite{KanePRL,DinePRL}.

Supergravity is specified in terms of two functions: the K\"ahler function
$G=K+\ln|W|^2$, where $K$ is the K\"ahler potential and $W$ the superpotential;
and the gauge kinetic function $f$. From these one can obtain the
supersymmetry-breaking scalar ($\widetilde m_i$) and gaugino masses and scalar
interactions ($A_i$) at the Planck scale, in terms of the gravitino mass
$m_{3/2}=e^{G/2}=e^{K/2}\,|W|$. In all known instances one obtains $\widetilde
m_i\sim c_i\,m_{3/2}$, where the $c_i={\cal O}(1)$ coefficients depend on the
specific functional form of $K$ and its field dependences. A similar result is
obtained for the trilinear scalar couplings, where typically $W$ also enters.
These model dependencies and potential non-universalities are however washed
out in the case of a light gravitino, where one effectively obtains
$m_0=A_0=0$. The bilinear scalar coupling $B_0$ may also vanish along these
lines, although this is not a general result as it may depend on the assumed
origin of the superpotential Higgs mixing parameter $\mu$. A gravitino mass of
suitable size can be easily obtained via gaugino condensation in the hidden
sector at the scale $\Lambda\sim10^k\,{\rm GeV}$: $m_{3/2}\sim
|W|\sim\Lambda^3/M^2\sim10^{3(k-9)}\,{\rm eV}$, where $M\approx10^{18}\,{\rm
GeV}$ is the appropriate gravitational scale. Cosmological and laboratory
constraints require
$10^{-5}\,{\rm eV}\raisebox{-4pt}{$\,\stackrel{\textstyle{<}}{\sim}\,$}
m_{3/2}\raisebox{-4pt}{$\,\stackrel{\textstyle{<}}{\sim}\,$} 10^3\,{\rm eV}$,
which entails $10^7\,{\rm GeV}
\raisebox{-4pt}{$\,\stackrel{\textstyle{<}}{\sim}\,$}\Lambda
\raisebox{-4pt}{$\,\stackrel{\textstyle{<}}{\sim}\,$}10^{10}\,{\rm GeV}$.
Condensation scales in this range are obtained for hidden gauge groups like
SU(3) and SU(4) with light hidden matter fields.

The gaugino masses depend on the gauge kinetic function ($f$), as follows
\begin{equation}
m_{1/2}=m_{3/2}\left({\partial_z f\over 2{\rm Re} f}\right)
\left({\partial_z G \over \partial_{z z^*} G}\right)\ ,
\label{eq:formula}
\end{equation}
where $z$ represents the hidden sector (moduli) fields in the model, and the
gaugino mass universality at the Planck scale is insured by a gauge-group
independent choice for $f$. The usual expressions for $f$ give $m_{1/2}\sim
m_{3/2}$. This undesirable result in the light gravitino scenario can be
avoided by considering the non-minimal choice $f\sim e^{-A z^q}$, where $A,q$
are constants. Assuming the standard no-scale expression $G=-3\ln(z+z^*)$,
one can then readily show that \cite{EEN1}
\begin{equation}
m_{1/2}\sim \left({m_{3/2}\over M}\right)^{1-{2\over3}q} M\ .
\label{eq:result}
\end{equation}
The phenomenological requirement of $m_{1/2}\sim10^2\,{\rm GeV}$ then implies
${3\over4}\raisebox{-4pt}{$\,\stackrel{\textstyle{>}}{\sim}\,$} q
\raisebox{-4pt}{$\,\stackrel{\textstyle{>}}{\sim}\,$}
{1\over2}$ for $10^{-5}\,{\rm eV}
\raisebox{-4pt}{$\,\stackrel{\textstyle{<}}{\sim}\,$}
m_{3/2}\raisebox{-4pt}{$\,\stackrel{\textstyle{<}}{\sim}\,$}10^3\,{\rm eV}$.
Note that $q={3\over4}$ gives the relation $m_{3/2}\sim
m^2_{1/2}/M\sim10^{-5}\,{\rm eV}$, which was obtained very early on in
Ref.~\cite{BFN} from the perspective of hierarchical supersymmetry breaking in
extended N=8 supergravity. The recent theoretical impetus for supersymmetric
M-theory in 11 dimensions may also lend support to this result, as N=1 in D=11
corresponds to N=8 in D=4.

Enforcing the constraints from radiative electroweak symmetry breaking in the
usual manner we obtain the low-energy supersymmetric spectrum in terms of two
parameters: $\tan\beta$ and $m_{1/2}$ (as well as the sign of $\mu$ and our
choice of $m_t=175\,{\rm GeV}$). Enforcing also $B_0=0$ allows one to solve for
$\tan\beta$ in terms of $m_{1/2}$, giving our one-parameter model. (The sign of
$\mu$ gets also determined in this process.) In practice allowing $B_0$ to
`float' does not qualitatively change the model predictions, although it dulls
them somewhat. These one- and two-parameter supergravity models have been
effectively considered before \cite{One,Easpects} without explicit mention of
what the gravitino mass was. The restriction $m_{3/2}\ll m_{1/2}$ does not
alter the spectra, but it changes the experimental signals that must now always
contain hard photons from $\chi^0_1$ decay. To be consistent with
our soft-supersymmetry-breaking assumptions at the Planck scale, we have
started the renormalization-group evolution at that scale (as in the case of
string models). As is well known, unification of the gauge couplings in this
type of scenario requires the introduction of intermediate-scale particles,
which we have implemented as described in Refs.~\cite{One,Easpects}.

\begin{figure}[p]
\vspace{6in}
\includegraphics{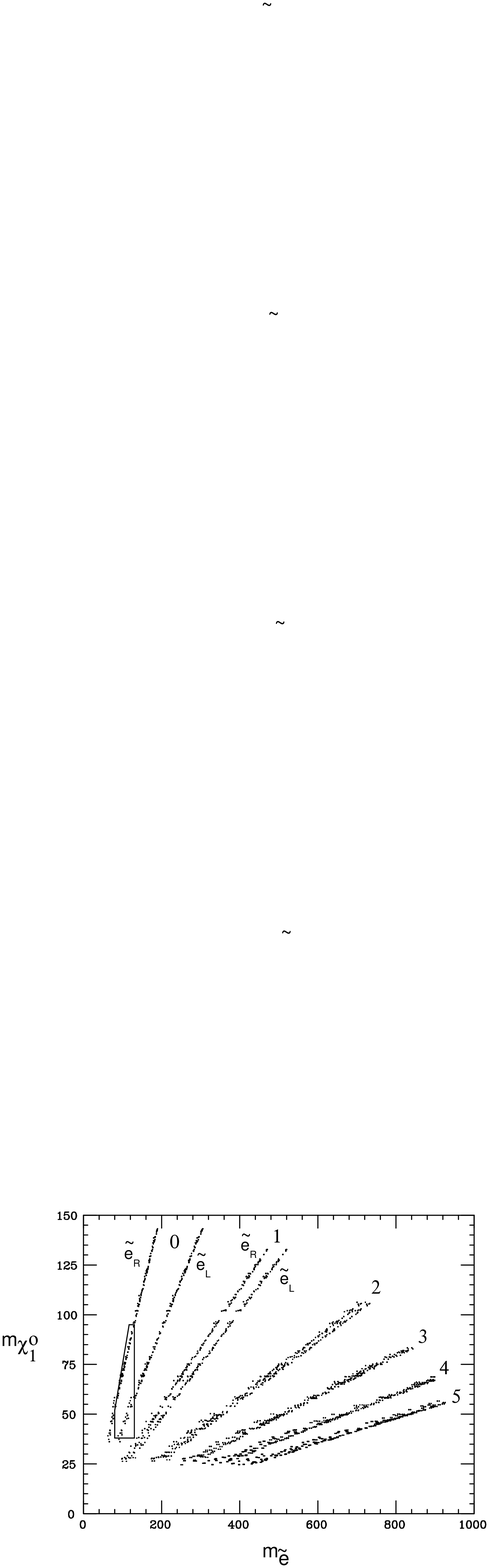}
\caption{Calculated distribution of selectron ($\tilde e$) and lightest
neutralino ($\chi^0_1$) masses (in GeV) in generic supergravity models for
fixed values of the ratio $\xi_0=m_0/m_{1/2}=0,1,2,3,4,5$; and varying values
of $\{m_{1/2}, \tan\beta,A_0\}$. The area within the polygon is consistent with
the kinematics of the CDF $ee\gamma\gamma$ event. The branches for $\tilde e_R$
and $\tilde e_L$ are only distinguishable for $\xi_0=0,1$. Our model predicts
$\xi_0=0$.}
\label{fig:lspsel}
\end{figure}

The question may arise of why of all possible unified supergravity models
described in terms of four parameters ($m_{1/2},m_0,A_0,\tan\beta$) should one
pay particular attention to our one-parameter model.
To gain some insight into this question we have generated 10,000 different
random four-parameter sets of this kind, and in each case determined the
low-energy spectrum, in particular the $\chi^0_1$ and $\tilde e_{R,L}$ masses.
The known kinematics of the $ee\gamma\gamma$ event in the light gravitino
scenario allow one to delineate an allowed region in the
$(m_{\tilde e},m_{\chi^0_1})$ plane \cite{KanePRL}. Figure~\ref{fig:lspsel}
shows the distribution of models in this space, with the preferred region
delineated by the polygon. For clarity, in the figure we restrict the choices
of $\xi_0=m_0/m_{1/2}$ to the integer values shown ({\em i.e.}, $0\to5$), with
the other three parameters allowed to vary at random. (The branches for $\tilde
e_R$ and $\tilde e_L$ are only distinguishable for  $\xi_0=0,1$.) This figure
illustrates the fraction of the generic supergravity parameter space that is
consistent with the kinematics of the CDF event. Moreover, our model prediction
of $\xi_0=0$ clearly falls within the allowed region for both $\tilde e_R$ and
$\tilde e_L$, whereas $\xi_0\ge1$ is not allowed.

\begin{figure}[p]
\vspace{6in}
\includegraphics{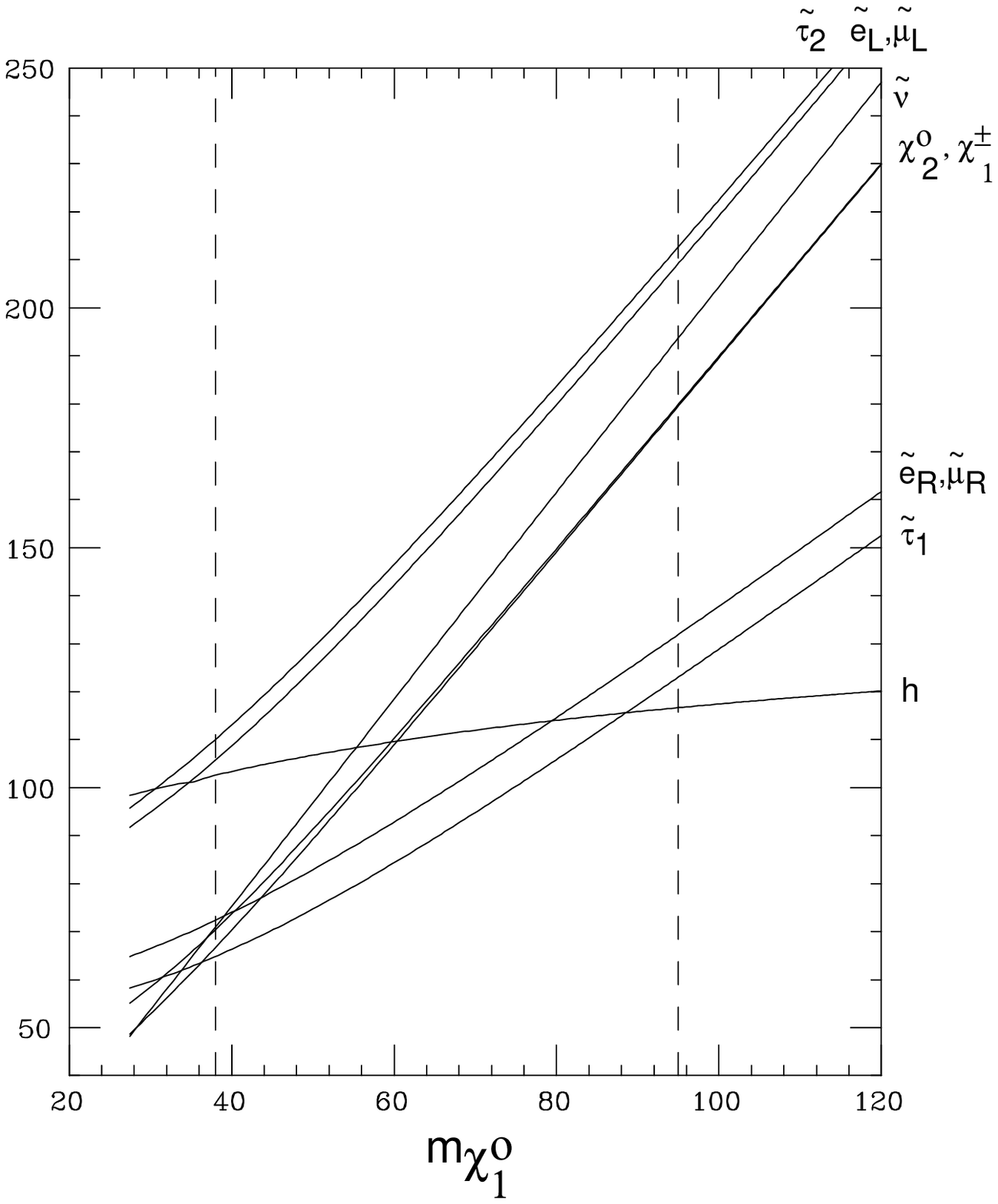}
\caption{The lighter members of the spectrum of our one-parameter model
versus the lightest neutralino mass. The vertical dashed lines delimit
the experimentally preferred region. All masses in GeV.}
\label{fig:light}
\end{figure}

\begin{figure}[p]
\vspace{6in}
\includegraphics{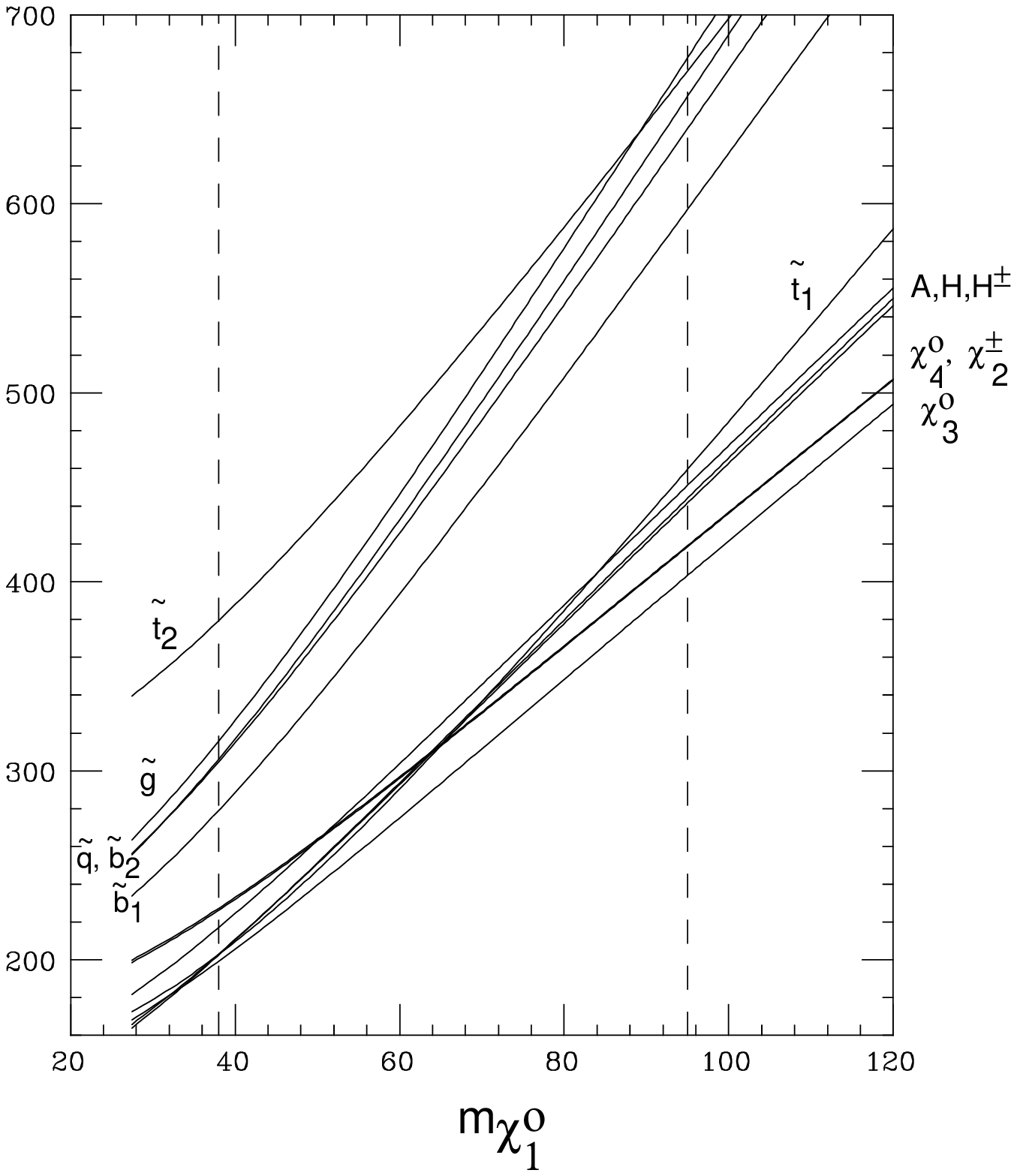}
\caption{The heavier members of the spectrum of our one-parameter model
versus the lightest neutralino mass. The vertical dashed lines delimit
the experimentally preferred region. All masses in GeV.}
\label{fig:heavy}
\end{figure}

We now turn to the phenomenological consequences of our one-parameter model.
The spectrum as a function of the lightest neutralino mass is given in
Fig.~\ref{fig:light} for the lighter particles (sleptons, lightest higgs,
lighter neutralinos and charginos) and in Fig.~\ref{fig:heavy} for the
heavier particles (gluino, squarks, heavy higgses, heavier neutralinos and
charginos). In addition we have the result $\tan\beta\approx8-10$. These
figures show that the lightest neutralino (which is mostly bino) is always
the next-to-lightest supersymmetric particle (NSLP), followed by the
right-handed sleptons ($\tilde e_R,\tilde\mu_R,\tilde\tau_1$), the lighter
neutralino/chargino ($\chi^0_2,\chi^\pm_1$), the sneutrino ($\tilde\nu$), and
the left-handed sleptons ($\tilde e_L,\tilde\mu_L,\tilde\tau_2$). (The order of
the second and third elements is reversed for very light neutralino masses.)
Note the splitting between the selectron/smuon masses and the stau mass due to
the non-negligible value of the $\lambda_\tau$ Yukawa coupling. The lightest
higgs boson crosses all sparticle lines with $m_h=(100-120)\,{\rm GeV}$. Also
notable is that the average squark mass is slightly below the gluino mass and
the lightest top-squark ($\tilde t_1$) is somewhat lighter than these.
The dominant decay of the lightest neutralino is via
$\chi^0_1\to\gamma\widetilde G$, which will proceed without suppression in the
experimentally preferred range $m_{\chi^0_1}\approx(38-95)\,{\rm GeV}$,
requiring only
$m_{3/2}\raisebox{-4pt}{$\,\stackrel{\textstyle{<}}{\sim}\,$}250\,{\rm eV}$ for
it to likely occur within the CDF (or any other such) detector \cite{KanePRL}.

\begin{figure}[p]
\vspace{6in}
\includegraphics{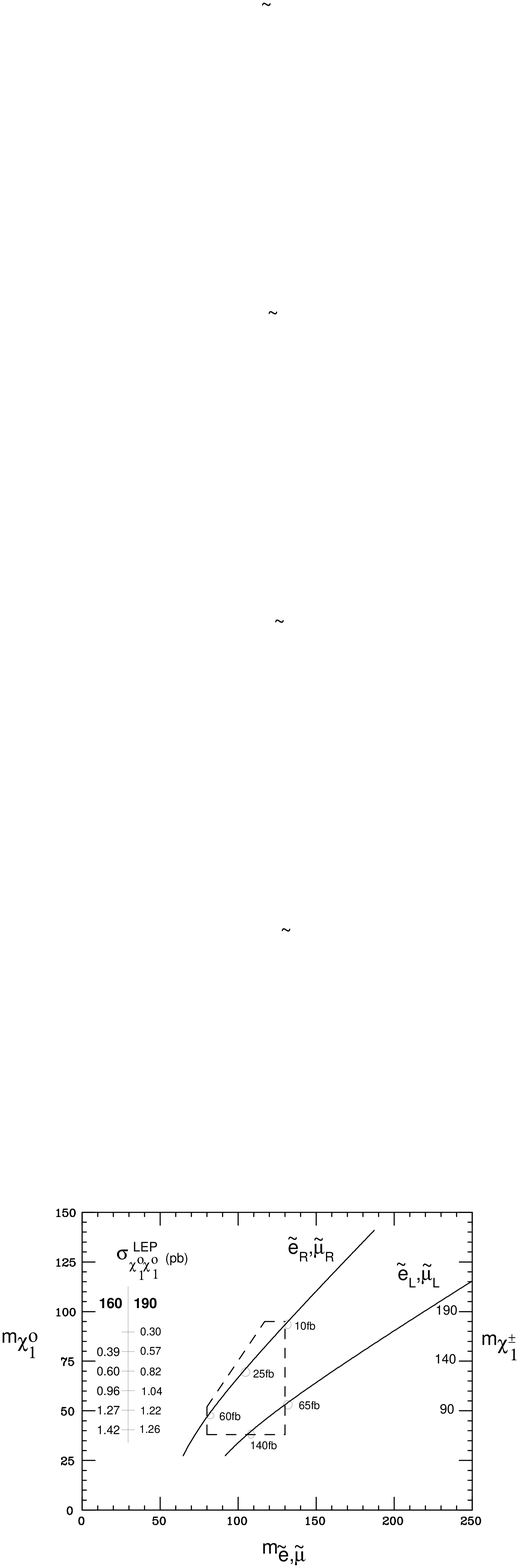}
\caption{The correlated predictions for the lightest neutralino mass
($m_{\chi^0_1}$) versus the selectron (or smuon) mass ($m_{\tilde
e,\tilde\mu}$) in our one-parameter model. The corresponding values of the
lightest chargino mass ($m_{\chi^\pm_1}$) are shown on the right axis.
(All masses in GeV.) The area within the polygon is consistent with the
kinematics of the CDF $ee\gamma\gamma$ event. The values at the marked points
along the lines indicate the cross section for the corresponding slepton
pair-production at the Tevatron. Also shown are the cross sections for
$e^+e^-\to\chi^0_1\chi^0_1\to\gamma\gamma+E_T\hskip-13pt/\quad$ production at
LEP160 and 190, as a function of $m_{\chi^0_1}$.}
\label{fig:B0}
\end{figure}

In Fig.~\ref{fig:B0} we show the correlated values of the lightest neutralino
mass versus the selectron (or smuon) mass. The lightest chargino mass (which
obeys $m_{\chi^\pm_1}\approx m_{\chi^0_2}$) is also shown in the figure. As the
figure shows, the experimentally preferred region (polygon) overlaps our model
predictions significantly for both $\tilde e_R$ and $\tilde e_L$. Moreover, the
cross section for pair-production of such particles at the Tevatron, as
indicated for a few points in the figure, shows that indeed only a few events
should have been produced in $0.1\,{\rm fb}^{-1}$ of data so far. Note also
that in the (preferred) case of $\tilde e_R$, the real constraint on our
one-parameter spectrum is on the selectron mass, the constraint on the
neutralino mass follows automatically. Our model is thus consistent with the
kinematics and dynamics of the CDF event.

LEP~1 constraints on our model are satisfied by construction. Because the
allowed region implies $m_{\tilde e}>80\,{\rm GeV}$ and $m_{\chi^\pm_1}\approx
m_{\chi^0_2}>70\,{\rm GeV}$ (see Fig.~\ref{fig:B0}), LEP~1.5
($\sqrt{s}=(130-136)\,{\rm GeV}$) was only sensitive to $\chi^0_1\chi^0_1$
production. A recent analyis by the OPAL Collaboration \cite{OPAL} of acoplanar
photon pairs at LEP~1.5 puts a 95\%CL upper limit on such cross section of
2~pb. We find $\sigma(e^+e^-\to\chi^0_1\chi^0_1)<1.6\,{\rm pb}$, and thus
LEP~1.5 imposes no new constraints on our model. Constraints from the Tevatron
are harder to determine. The lower bound on the selectron masses from
Fig.~\ref{fig:B0} implies $m_{\tilde g}>365\,{\rm GeV}$, $m_{\tilde
q}>350\,{\rm GeV}$, $m_{\tilde t_1}>235\,{\rm GeV}$ (see Fig.~\ref{fig:heavy}),
all of which automatically satisfy present Tevatron limits. However,
neutralino/chargino production via $p\bar
p\to\chi^+_1\chi^-_1,\chi^0_2\chi^\pm_1$ have larger rates leading to
$\gamma\gamma$+$n\ell$+$mj$+$E_T\hskip-13pt/\quad$ signals that might have been
detected. ($\chi^0_i\chi^0_j$ production is also kinematically accessible but
negligible because of the dominant gaugino nature of $\chi^0_{1,2}$.)
Ref.~\cite{Kane} estimates that the apparent non-observation of such processes
at the Tevatron with $100\,{\rm pb}^{-1}$ of data requires
$m_{\chi^\pm_1}>125\,{\rm GeV}$. Taken at face value, this constraint would
eliminate the lighter half of the allowed parameter space, marked by the
central point on the $\tilde\ell_R$ curve in Fig.~\ref{fig:B0}, and single
out $\tilde e_R$ as the only possible explanation for the event ({\em i.e.},
$m_{\chi^\pm_1}>125\,{\rm GeV}$ implies $m_{\tilde e_L}>155\,{\rm GeV}$).

The presumed lower bound on the chargino mass from Tevatron searches makes
$\tilde\ell^+\tilde\ell^-,\chi^+_1\chi^-_1,\chi^0_1\chi^0_2,\chi^0_2\chi^0_2$,
and higgs production at LEP160 or 190 kinematically disallowed. The only
accessible channel is
$e^+e^-\to\chi^0_1\chi^0_1\to\gamma\gamma\widetilde G\widetilde G$, which for
our bino-like neutralino proceeds dominantly via $t$-channel $\tilde e_R$
exchange. The cross sections for this process at LEP~160 and 190 are shown in
Fig.~\ref{fig:B0} as a function of $m_{\chi^0_1}$. It is not clear what
sensitivity will LEP160 have for such signal with the expected $10\,{\rm
pb}^{-1}$ of data. At LEP190 with $500\,{\rm pb}^{-1}$ of data, a detailed
study \cite{Kane} shows that it should be able to probe {\em all} of the
preferred range: $m_{\chi^0_1}\approx(38-95)\,{\rm GeV}$.

Any light gravitino scenario can be distinguished from the neutralino LSP
scenario, for instance, by the nature of the photon spectrum in, {\em e.g.},
$e^+e^-\to\chi^0_1\chi^0_1\to\gamma\gamma\widetilde G\widetilde G$
versus $e^+e^-\to\chi^0_2\chi^0_2\to\gamma\gamma\chi^0_1\chi^0_1$
\cite{DinePRL}. Our supergravity light gravitino model can be further
distinguished from the gauge-mediated models by the differing predicted
spectra, although the gauge-mediated ones depend on the unknown nature of their
`messenger sector'.

Concerning the cosmological aspects of our model, as is well known, for
$m_{3/2}\sim1\,{\rm KeV}$ the relic gravitinos constitute a form of `warm' dark
matter with a behavior similar to that of cold dark matter. The non-thermal
gravitinos from $\chi^0_1$ decay do not disturb big bang nucleosynthesis, and
may constitute a form of hot dark matter \cite{BMY}, although with small
abundance. Other forms of dark matter, such as metastable hidden sector
matter fields (cryptons) \cite{cryptons} and a cosmological constant
\cite{Age}, may need to be considered as well.

If the CDF event is really of supersymmetric origin, because of its peculiar
properties it would not only establish the existence of supersymmetry but also
provide strong clues as to the origin of supersymmetry breaking. Our no-scale
supergravity model with a light gravitino has strong roots in strings, even
in the modern era of extended supergravity and M-theory \cite{BD}.

\section*{Acknowledgments}
J. L. would like to thank Geary Eppley and Teruki Kamon for useful discussions.
The work of J.~L. has been supported in part by DOE grant DE-FG05-93-ER-40717.
The work of D.V.N. has been supported in part by DOE grant DE-FG05-91-ER-40633.

\newpage
\clearpage

\clearpage

\clearpage

\clearpage


\begin{thebibliography}{99}
\bibitem{EFL}See, {\em e.g.}, J. Ellis, G. Fogli, and E. Lisi, Z. Phys. C
{\bf69} (1996) 627.
\bibitem{EKN}J. Ellis, S. Kelley, and D.~V.~Nanopoulos, Phys. Lett. B {\bf249}
(1990) 441 and Phys. Lett. B {\bf260} (1991) 131; P. Langacker and M. Luo,
Phys. Rev. D {\bf44} (1991) 817; U. Amaldi, W. de Boer, and H. F\"urstenau,
Phys. Lett. B {\bf260} (1991) 447; F. Anselmo, L. Cifarelli, A. Peterman, and
A. Zichichi, Nuovo Cim. {\bf104A} (1991) 1817.
\bibitem{KanePRL}S. Ambrosanio, G. Kane, G. Kribs, S. Martin, and S. Mrenna,
Phys. Rev. Lett. {\bf76} (1996) 3498.
\bibitem{DinePRL}S. Dimopoulos, M. Dine, S. Raby, and S.
Thomas, Phys. Rev. Lett. {\bf76} (1996) 3502.
\bibitem{Park}S. Park, in Proceedings of the 10th Topical Workshop on
Proton-Antiproton Collider Physics, Fermilab, 1995, edited by R. Raja and J.
Yoh (AIP, New York, 1995), p. 62.
\bibitem{Kane}S. Ambrosanio, G. Kane, G. Kribs, S. Martin, and S. Mrenna,
hep-ph/9605398.
\bibitem{Dine}D. Stump, M. Wiest, and C.-P.~Yuan, hep-ph/9601362;
S. Dimopoulos, S. Thomas, and J. Wells, hep-ph/9604452;
K. Babu, C. Kolda, and F. Wilczek, hep-ph/9605408.
\bibitem{no-scale}E. Cremmer, S. Ferrara, C. Kounnas, and D.~V.~Nanopoulos,
Phys. Lett. B {\bf133} (1983) 61; J. Ellis, C. Kounnas, and D.~V.~Nanopoulos,
Nucl. Phys. B {\bf241} (1984) 406 and B {\bf247} (1984) 373. For a review see
A. Lahanas and D.~V.~Nanopoulos, Phys. Rep.~{\bf145} (1987)~1.
\bibitem{Lahanas} J. Ellis, A. Lahanas, D.~V.~Nanopoulos, and K. Tamvakis,
Phys. Lett. B {\bf134} (1984) 429.
\bibitem{EEN1}J. Ellis, K. Enqvist, and D.~Nanopoulos, Phys. Lett. B {\bf 147}
(1984) 99.
\bibitem{EEN2}J. Ellis, K. Enqvist, and D.~Nanopoulos, Phys. Lett. B {\bf 151}
(1985) 357.
\bibitem{LN}J.~L.~Lopez and D.~V.~Nanopoulos, hep-ph/9412332 (Int. J. Mod.
Phys. A).
\bibitem{cosmo}H. Pagels and J. Primack, Phys. Rev. Lett. {\bf48} (1982) 223;
S. Weinberg, Phys. Rev. Lett. {\bf48} (1982) 1303; J. Ellis, J. Kim, and
D.~V.~Nanopoulos, Phys. Lett. B {\bf145} (1984) 181.
\bibitem{BFN}R. Barbieri, S. Ferrara, and D.~V.~Nanopoulos, Phys. Lett. B
{\bf107} (1981) 275.
\bibitem{One}J.~L.~Lopez, D.~V.~Nanopoulos, and A. Zichichi, Int. J. Mod. Phys.
A{\bf 10} (1995) 4241.
\bibitem{Easpects}J.~L.~Lopez, D.~V.~Nanopoulos, G. Park, X. Wang, and A.
Zichichi, Phys. Rev. D {\bf50} (1994) 2164.
\bibitem{OPAL}G. Alexander, {\em et. al.} (OPAL Collaboration),
CERN-PPE/96-039.
\bibitem{BMY}S. Borgani, A. Masiero, and M. Yamaguchi, hep-ph/9605222.
\bibitem{cryptons}J. Ellis, J.~L.~Lopez, and D.~V.~Nanopoulos,
Phys. Lett. B {\bf247} (1990) 257.
\bibitem{Age}J.~L.~Lopez and D.~V.~Nanopoulos, Mod. Phys. Lett. A{\bf9} (1994)
2755; A{\bf11} (1996)~1.
\bibitem{BD}See {\em e.g.}, T. Banks and M. Dine, hep-th/9605136.
\end{thebibliography}
\end{document}